\documentclass[twocolumn,twoside]{revtex4}
\usepackage{graphicx}
\usepackage{fancyhdr}
\usepackage{xcolor}
\graphicspath{{../}{../../../Shingaku_iga/figures/} {../../../figs/}}

\begin{document}

\title{Precision Measurements of the PMNS Parameters with T2K Data}

%

\author{Ali Ajmi\\
on behalf of the {\bf T2K Collaboration}}
\affiliation{University of Winnipeg, Canada, R3B2E9}

\begin{abstract}
 
 T2K is a long baseline neutrino experiment which exploits a neutrino and antineutrino beam at JPARC to
perform precision measurements of neutrino oscillation parameters $\Delta {\rm m}^2_{\rm 32}$, $\sin ^2 \theta_{23}$ 
(besides the CP-violating phase $\delta_{\rm CP}$). The latest results  for the measurement of PMNS
parameters in the disappearance mode are presented here, highlighting the main systematic uncertainties limiting the precision. The
future strategy to improve the precision on the measurement of PMNS parameters are also discussed.
\end{abstract}

\maketitle

\thispagestyle{fancy}


\section{Introduction}
The T2K (Tokai to Kamioka)  \cite{t2k} is a long baseline neutrino experiment, located in Japan, sending muon-(anti)neutrino beam, produced from 30 GeV protons at the Japan Proton Accelerator Research Complex (JPARC) at Tokai to Kamioka, along the 295 km baseline. The 2.5$^\circ$off-axis beam  peaks around 0.6 GeV.  The near detectors for this experiment are located 280m downstream from the source at JPARC in Tokai, and the  water-cherenkov detector, Super-Kamikande at Kamioka plays the role of the far detector for studying neutrino oscillations. 

The basics of neutrino oscillations, the T2K experiment and the results from the appearance mode at T2K are described in \cite{Joe}. The latest results  { \cite{nature, pdun}} from the disappearance mode are discussed here. The aim is to measure the value of the parameters $\Delta {\rm m}^2_{32}$ and the $\theta_{23}$ with maximum possible precision. This requires one to ensure minimum systematic uncertainties in the experiment, which is addressed through a number of contributing factors, like the flux, the neutrino interaction models, and the constraints from the near detectors, to name a few.

These factors are discussed in the following sections, along with the new improvements being worked upon at T2K for more precise oscillation analyses in the near future.
 
\section{Data for the presented results}

The T2K experiment has been taking data since 2010 and has
accumulated a total of 1.99$\times$10$^{21}$ protons-on-target (POT) in $\nu_\mu$ mode and 1.65$\times$10$^{21}$ POT in $\bar{\nu}_\mu$ mode until 2020. 
 { The near detector used 1.15$\times$10$^{21}$ POT in neutrino mode and 0.83$\times$10$^{21}$ POT in antineutrino mode for the presented analysis results. The details of data used at the far detector are mentioned in \cite{Joe}.}
The oscillation analysis procedure followed at T2K is explained in \cite{Joe, pdun}.

\section{Flux-prediction uncertainties}

\begin{figure}[h]
\centering
\includegraphics[width=80mm]{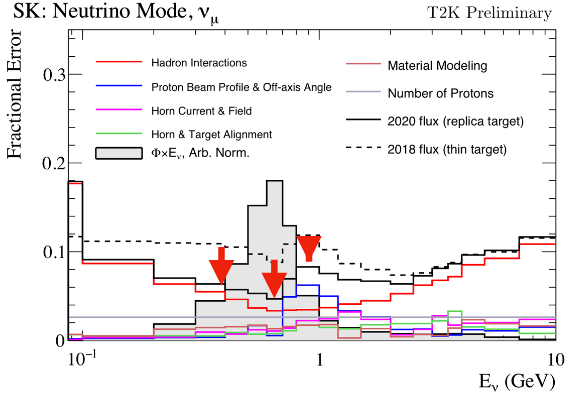}
\caption{Flux prediction uncertainties vs neutrino energy in the neutrino mode for T2K, at the far detector site.  {The arrows highlight the reduction in the uncertainties from the earlier thin target case (dotted line) to the replica target case (bold line).}} \label{fig1}
\end{figure}

The neutrino flux for the T2K is simulated with the FLUKA simulation package \cite{flux},
and then tuned with external data constraints from the NA61/SHINE   hadron production
experiment at CERN.
Earlier, a thin graphite target was used for this purspose, but changing it to the   target configuration replica of that of T2K \cite{replica},  the systematic uncertainties on the flux reduced from 8\% to 5\%, as shown in Fig.~\ref{fig1}. The current analysis uses this uncertainty.

Further improvements can be achieved by using the NA61/SHINE 2010 {data \cite{shine},} which adds kaons and protons yields along with increased statistics, reducing the uncertainties further to $\sim$4\%. Results with this will be presented soon.

\section{The Near Detector (ND) Constraints}

The detector along the axis of the beam, built out of scintillator-iron layers (INGRID) monitors the incident neutrino beam and its stability.
The off-axis near detector suite  {(ND280)} comprises of a magnet, two fine-grained detectors (FGD1: scintillators, FGD2: scintillator-water layers) in between three time projection chambers (TPCs) to act as the tracker, the Electromagnetic Calorimeter (scintillator-lead), the Pi-Zero detector (scintillator-water bags) and the Side-Muon Range Detector (scintillator plates). The magnetic field of 0.2 T helps in charge identification of the particles.

The dominant interaction type at the energy range for T2K, is the charged-current quasi-elastic
reaction (CCQE). 
Other charged-current interactions are also present, like mainly the  resonant pion production (CCRES), 
and the deep
inelastic scattering (CCDIS) channels.

The events in the near detector are classified according to the topology based on the reconstructed pion multiplicity. 

\subsection{Neutrino Interaction Models:}

One of the largest and most complex systematic uncertainty in the interpretation of the data for the neutrino  oscillation measurement is due to the modelling of neutrino-nucleus interactions. The T2K collaboration developed a new version of the model of neutrino-nucleus interactions based on the Spectral Function approach and with refined tuning of the nucleons removal energy. 

The T2K Collaboration has been improving the NEUT neutrino event generator  iteratively with every analysis. 
  Significant updates have been applied on the recent NEUT 5.4.0 model \cite{neut}. 
  A tuned Benhar Spectral Functiforon \cite{benhar} to describe CCQE interactions is being used now instead of the earlier used Relativistic Fermi Gas Model with Random Phase Approximations. 
   The Shell model is built largely from the electron-scattering data \cite{elecdata}, with the nuclear ground states better defined and the outgoing nucleon kinematics better predicted. This model is being used for the current results.  

Further modifications are underway, one of which is the $|q_3|$-dependent removal energy treatment from comparing NEUT
to electron scattering data (where, q$_3$ is the three-momentum transfer in nuclear models). All these modifications will make NEUT a more robust neutrino interaction model and amount to reduced uncertainties from the interaction modelling.

\subsection{Events Selections} 
 
 Events at the ND280 are selected with mainly three types of topologies: the charged current no-pion (CC0pi) events,  the CC1pi (events containing only one pion track besides the lepton) and the CC-others (events containing multiple tracks), which are enriched in the CCQE, CCRES, CCDIS channels respectively. A total of 18 such samples (FGD1, FGD2, both modes and wrong-sign component for the anti-neutrino mode) are used for this analysis, with larger data statistics in comparison to earlier results. 

Further improvements in the selection processes are being made for future analyses, by implementing proton/gamma tagging methods \cite{kamil}. The newer criteria are based on the energy and charge depositions, the likelihood ratios and the electro-magnetic shower topologies, all resulting to an increase in the purity of the samples by 5-10\%. Newer samples with multi-ring topologies at SuperK are also being worked upon \cite{lsm}. 

\subsection{Near Detector Fits}
An extended binned likelihood fit to the
ND sample as a function of muon kinematics is made to constrain the predicted number of events. The resulting post-fit ND events distributions match the observed data well,  {with a prior model p-value of 74\%.} The fitting introduces anti-correlation between the flux and the cross section model parameters. 

\begin{figure}[h]
\centering
\includegraphics[width=80mm]{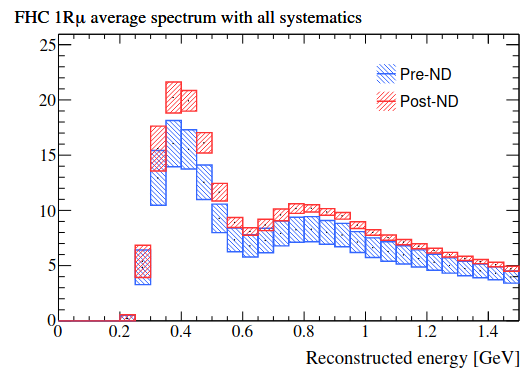}
\caption{Predicted flux at the FD site with the systematic uncertainties shown before and after the ND fits, for the Forward Horn Current (FHC) or the neutrino mode. } \label{fig2}
\end{figure}

The flux prediction  {of the muon neutrinos} at the Far-Detector (FD) is made thereafter, and the ND-fit reduces the uncertainty to a great extent. The systematic uncertainties at SK are hence 3.0 (4.0)\% in $\nu$-mode ($\bar\nu$-mode), as can be seen in Fig.\ref{fig2}.
The uncertainties in the different samples are listed in  {Table~\ref{tab1}.}

\begin{table}[h]
\begin{center}
\caption{Systematic Uncertainties at the far detector before and after using flux, cross section model,  ND constraints.  (d.e. stands for `decay electron') }\label{tab1}
\begin{tabular}{|l|c|c|}
\hline \textbf{SuperK Sample} & \textbf{Pre-ND fit error} & \textbf{Post-ND fit error} \\
\hline $\nu_\mu$ 1R$_\mu$ & 11.1\% & 3.0\% \\
 $\bar{\nu}_\mu$ 1R$_\mu$  & 11.3\% & 4.0\% \\
\hline $\nu_e$ 1R$_e$  & 13.0\% & 4.7\% \\
 $\bar{\nu}_e$ 1R$_e$& 12.1\% & 4.9\% \\
\hline $\nu_e$ 1R$_e$ 1 d.e. & 18.7\% & 14.3\% \\
\hline
\end{tabular}
\label{tab1}
\end{center}
\end{table}

\section{Results presented}

\begin{figure}[h]
\centering
\includegraphics[width=80mm]{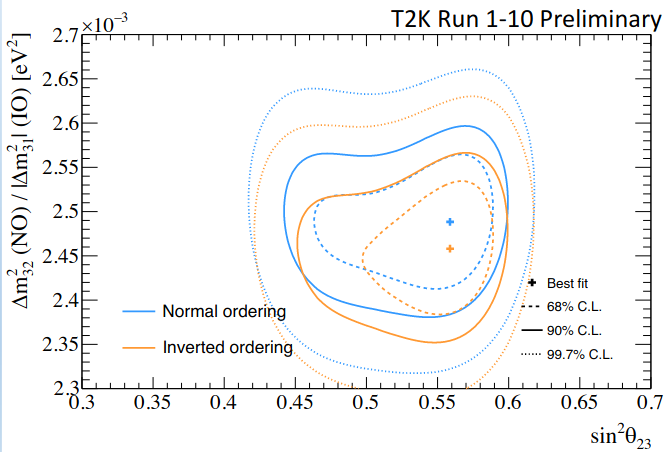}
\caption{Best fit results from T2K (reactor constraints included) for the $\Delta {\rm m}^2_{\rm 32}$ and $\sin ^2 \theta_{23}$. } \label{fig3}
\end{figure}

The oscillation analysis was done using five SK samples for T2K Run 1-10, as elaborated in \cite{Joe}. The upper octant is slightly preferred with a probability of 77.1\%. Similar results are also noticed by the Frequentist approach, as shown in Fig.~\ref{fig3}.  {The ND-fits and the FD-fits are done consecutively in the Frequentist approach, while the ND and the FD fits are done simultaneously in the Bayesian approach.} The normal ordering of the neutrino mass hierarchy is preferred at 80.8\% probability, and is well in agreement with  the Frequentists confidence intervals. The Bayesian results are listed in Table~\ref{tab2} and~\ref{tab3}.

\begin{table}[h]
\begin{center}
\caption{Best fit results from T2K (with and without reactor constraints by the Bayesian method) for the   $\Delta {\rm m}^2_{\rm 32}$. }\label{tab2}
\begin{tabular}{|l|c|p{2cm}|c|}
\hline  & \textbf{Hierarchy} & \textbf{Most Probable Value} & \textbf{Range} \\
        &   & (in 10$^{-3}$ eV$^2$) &  (in 10$^{-3}$ eV$^2$) \\
\hline T2K only & Normal & 2.487   & [2.437, 2.537] \\
                & Inverted & 2.457 & [2.407, 2.507] \\
\hline T2K + reactor & Normal & 2.485   & [2.436, 2.536]  \\
                & Inverted & 2.457   & [2.406, 2.506] \\
\hline
\end{tabular}
\label{tab2}
\end{center}
\end{table}

\begin{table}[h]
\begin{center}
\caption{Best fit results from T2K (with and without reactor constraints by the Bayesian method) for the   $\sin ^2 \theta_{23}$. } \label{tab3}
\begin{tabular}{|l|c|p{1.9cm}|p{2.5cm}|}
\hline  & \textbf{Hierarchy} & \textbf{Most Probable Value} & \textbf{Range} \\
\hline T2K only & Normal & 0.471   & [0.452, 0.508] and [0.530, 0.568] \\
                & Inverted & 0.469 & [0.449, 0.508] and [0.531, 0.565] \\
                & Both & 0.471 & [0.451, 0.508] and [0.530, 0.567] \\
\hline T2K + reactor & Normal & 0.559   & [0.504, 0.583]  \\
                & Inverted & 0.560 & [0.519, 0.585]   \\
                & Both & 0.559 & [0.507, 0.584]   \\
\hline
\end{tabular}
\label{tab3}
\end{center}
\end{table}
%
%

\section{T2K Upgrade}

The T2K beam will be upgraded with increased power of $\sim$750kW in 2022 and an upgraded version  of the ND280 near detector is being assembled to exploit the increased statistics \cite{beamupgrade, t2kupgrade}. 

Two of the beam's magnetic focussing horns are being replaced at JPARC, the horn power supply is being upgraded to enable faster beam repetition rate, and efforts are going on to improve the cooling capability of the beam target.

Moreover in 2020, the
Super-Kamiokande detector was loaded with 0.01\% of Gadolinium  \cite{skupgrade}, enhancing its neutron tagging capabilities.
The T2K Run-11 data was taken with this configuration, the analysis of which is currently going on.

The near detector is undergoing a major upgrade. A Super-FGD comprising of $\sim$2 million scintillator cubes, and High Angle TPCs are being added for more phase space acceptance, and improved resolution in the particle kinematics, thus increasing the detection efficiency of protons, neutrons, lower momentum pions, and also aid in tagging decay electrons.

\section{Summary}

   Latest precise measurements of the neutrino oscillation parameters from the T2K experiment, using 3.8$\times 10^{21}$ POT  data have been presented here for the disappearance mode. A slight preference for Non-maximal mixing is observed. The data shows a preference for the upper octant  for $\theta_{23}$ with a 77.1\% probability. The normal ordering of the neutrino mass hierarchy is also preferred by the observed  data with 80.8\% probability.
   
   However, several efforts are being made to achieve improved sample selections, reduced predicted flux uncertainties, and more robust cross section models. These will result into more precise values of the measured oscillation parameters which will be presented to the physics community very soon.
   
%


\bigskip 

\end{document}